\documentclass[%
twocolumn,
 superscriptaddress,
 amsmath,amssymb,
 aps, physrev,
]{revtex4-2}

\usepackage{graphicx}
\usepackage[usenames,dvipsnames]{xcolor}
\usepackage{dcolumn}
\usepackage{bm}

\usepackage{siunitx}
\usepackage{upgreek}
\usepackage[colorlinks=true]{hyperref}  
\hypersetup{
    bookmarks=true,         
    unicode=false,          
    pdftoolbar=true,        
    pdfmenubar=true,        
    pdffitwindow=false,     
    pdfstartview={FitH},    
    pdftitle={OrbitalDiode},    
    pdfauthor={Perego},     
    pdfsubject={},   
    pdfcreator={},   
    pdfproducer={}, 
    pdfkeywords={} {} {}, 
    pdfnewwindow=true,      
    colorlinks=true,       
    linkcolor=blue, 
    citecolor=blue,        
    filecolor=magenta,      
    urlcolor=blue           
}

\begin{document}

\title{\textbf{In-plane and out-of-plane magnetic field driven Josephson diode effect \\ in magic-angle twisted four-layer graphene} 
}%

\author{Marta Perego}	
\email{mperego@phys.ethz.ch}
\affiliation{Laboratory for Solid State Physics, ETH Zurich,~CH-8093~Zurich, Switzerland}
\affiliation{Quantum Center, ETH Zurich,~CH-8093 Zurich, Switzerland}
\author{Sayan Banerjee}
\affiliation{Institute for Theoretical Physics III, University of Stuttgart, 70550 Stuttgart, Germany}
\author{Mohamad Abu El Hija}
\author{Clara Galante Agero}
\author{Alexandra Mestre-Tor\`a}
\author{Giovanni Zhang}
\affiliation{Laboratory for Solid State Physics, ETH Zurich,~CH-8093~Zurich, Switzerland}
\affiliation{Quantum Center, ETH Zurich,~CH-8093 Zurich, Switzerland}
\author{Takashi Taniguchi}
\affiliation{Research Center for Materials Nanoarchitectonics, National Institute for Materials Science,  1-1 Namiki, Tsukuba 305-0044, Japan}
\author{Kenji Watanabe}
\affiliation{Research Center for Electronic and Optical Materials, National Institute for Materials Science, 1-1 Namiki, Tsukuba 305-0044, Japan}
\author{Mathias S. Scheurer}
\affiliation{Institute for Theoretical Physics III, University of Stuttgart, 70550 Stuttgart, Germany}
\author{Thomas Ihn}
\author{Klaus Ensslin}
\affiliation{Laboratory for Solid State Physics, ETH Zurich,~CH-8093~Zurich, Switzerland}
\affiliation{Quantum Center, ETH Zurich,~CH-8093 Zurich, Switzerland}

\date{\today}

\begin{abstract}
\vspace{1em}
\centering{\textbf{Abstract}}\\[0.5em]
The superconducting diode effect offers a powerful probe into the fundamental symmetries of quantum materials. Recent studies on twisted graphene diodes have predominantly focused on bilayer or trilayer systems under out-of-plane magnetic fields.
Here, we demonstrate both out-of-plane and in-plane driven Josephson diode effects in a magic-angle twisted four-layer graphene junction, i.e., an even number of layers.
We observe the emergence of a diode effect at zero out-of-plane field, tuned by an increasing in-plane magnetic field. This result points to the presence of strong in-plane orbital coupling, which is highly sensitive to the specific layer parity of the structure.
Our findings provide experimental insights into the symmetry-breaking mechanisms of even-layer twisted graphene, establishing in-plane magnetic fields as a vital tool for unravelling their microscopic properties.

\end{abstract}

\maketitle

\section*{Introduction} 

Semiconductor-based diodes, with their electrical resistance being direction-dependent, are key components in electronic circuits. Their superconducting analogue, the superconducting diode, exhibits asymmetric critical supercurrents such that it is possible to obtain dissipationless current flow along only one direction \cite{nadeem2023superconducting,ma2025superconducting}. Conversely, driving the same current in the reverse direction drives the device into a dissipative state characterised by a finite voltage drop. Observing the superconducting diode effect (SDE) and studying its parameter dependence provides crucial information on the possible pairing mechanisms and intrinsic symmetries of the superconductor \cite{Alexreview}, while paving the way for the development of dissipationless rectification and memory elements \cite{ando2020observation, nadeem2023superconducting, ma2025superconducting,Alexreview}.

Twisted multilayer graphene represents a novel class of superconductors whose phenomenological and microscopic properties are the subject of intense ongoing research \cite{cao2018unconventional,liu2022isospin,kim2022evidence,park2025experimental,kim2026resolving,birkbeck2025quantum, lee2025revealing, Kim2024, Oliver2024, portoles2024quasiparticle, mukherjee2025superconducting}. Therefore, exploring the SDE in these systems has the potential to provide crucial insights into the superconducting state and pairing mechanism. 
Importantly, different numbers $N$ of graphene layers and twist-angle configurations have been realized in experiments and shown to provide a plethora of ways to engineer two-dimensional electron liquids. Focusing, for instance, on alternating twist-angle configurations, the moiré superlattice system exhibits a horizontal mirror symmetry only for odd $N$.
Moreover, each $N$-layer configuration presents unique magic twist angles \cite{khalaf2019magic,carr2017twistronics}, phase diagrams, and varying degrees of displacement field tunability, alongside different superconducting critical temperatures, fields, supercurrents \cite{cao2021pauli,park2022robust,zhang2022promotion,burg2022emergence, hao2021electric}, and Pauli limit violation \cite{cao2021pauli, park2022robust, talantsev2022compliance}.
Most experimental works on the SDE have been carried out in twisted bilayer graphene \cite{DimasFirstDiode, diez_second, RothsteinGateTunable} and twisted trilayer graphene \cite{Ronen2025, lin2022zero, zhang2024angle}, mainly through the characterisation of Josephson junctions (JJs). 
Being based on dissipationless processes, any realisation of the SDE, including the aforementioned Josephson diode effect (JDE), requires not only broken inversion and $C_{2z}$ rotational symmetry but also broken time-reversal symmetry.

At the moment, there is no universally agreed upon mechanism for symmetry breaking across different graphene-based samples displaying the SDE.
Possible mechanisms range from spontaneous breaking of both of these symmetries already in the normal state \cite{lin2022zero,DimasFirstDiode,zhang2024angle,Ronen2025}, most notably via valley polarization \cite{lin2022zero,DimasFirstDiode,Scammell2022Mar,PhysRevLett.130.266003,Han2024a}, inversion symmetry breaking via sublattice polarization \cite{diez_second}, and large kinetic inductance combined with non-uniform supercurrents \cite{RothsteinGateTunable}.

While studies predominantly focused on diode effects driven by out-of-plane magnetic fields, the in-plane field response remains largely unexplored. Therefore, developing a clear understanding of its role for the SDE, in particular, and superconductivity in twisted multilayer graphene, in general, is important. Apart from the Zeeman field \cite{pal_josephson_2022}, there is also an orbital effect of an in-plane magnetic field, associated with Peierls phases in the interlayer hopping matrix elements \cite{antebi2022plane, mandal2023valley, margetis2024optical, vasilevskiy2026plane, vishwanathInplaneCoupling}. As a result of the weak spin-orbit coupling in graphene \cite{huertas2006spin,konschuh2010tight}, the SDE is expected to be primarily driven by the latter, orbital coupling, which will allow us to focus on its role for superconductivity.

Here, we explore both the out-of-plane and in-plane magnetic field dependence of a gate-defined JJ in magic-angle twisted four-layer graphene (MAT4G). Our transport measurements reveal the presence of a JDE for both out-of-plane and in-plane directions of the magnetic field. Strikingly, we observe the emergence of a JDE at zero out-of-plane magnetic field, driven by an increasing in-plane field. We attribute this behaviour to the interplay between spatial moiré inhomogeneity along the transverse direction of the junction and \textit{in-plane orbital coupling}, suggesting that this orbital effect plays a key role in even-layered alternating twisted graphene.

\section*{Results}
\subsection*{Device, material and tuning}
\begin{figure}
\includegraphics[width=\columnwidth]{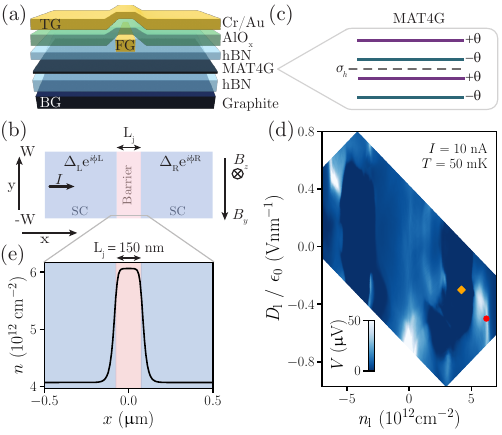}
\caption{(a) Illustrative structure of the device. The MAT4G is encapsulated using hexagonal boron nitride (hBN). At the bottom, a graphite layer serves as the bottom gate (BG), while the top gate (TG) and the finger gate (FG) are made of Cr/Au. (b) Josephson junction geometry with superconducting (SC) leads in blue and the resistive barrier area of length $L_\mathrm{j}$ in pink. The d.c.\ current $I$ flows along the $x$-axis, and the device has a width of $2W$ along the $y$-axis. The in-plane $B_y$ and out-of-plane $B_z$ magnetic field directions are indicated. The superconducting order parameters for the left and right leads are denoted as $\Delta_\mathrm{L,R}e^{i\phi_{L,R}}$. (c) Cross-section of MAT4G demonstrating an alternating twist sequence. The first and third graphene layers (purple) are twisted by a positive angle $+\theta$, while the second and fourth layers (green) are twisted by a negative angle $-\theta$. The dashed black line marks the horizontal midplane, highlighting the absence of horizontal mirror symmetry ($\sigma_h$). To preserve $\sigma_h$ symmetry, the twist angle sequence would instead need to be $+\theta, -\theta, -\theta, +\theta$. (d) Voltage $V$ measured at constant $I=\SI{10}{nA}$ and temperature $T=\SI{50}{mK}$ as a function of the leads' density $n_\textnormal{l}$ and displacement field $D_\textnormal{l}$. The orange rhombus indicates the leads tuning of the junction, while the red dot indicates the barrier tuning. Reproduced from Fig.~1 in Ref.~\cite{perego2024experimental}. (e) Electrostatic simulation profile of the carrier density as a function of position $x$ across the device, showing a close-up view around the junction region to highlight the extracted $L_\mathrm{j}$.
}
\label{Fig1} 
\end{figure}

Our device, which has previously been studied in \cite{perego2024experimental,perego2026pearl,Perego2026}, consists of MAT4G encapsulated in hexagonal boron nitride (hBN). A schematic illustration of the heterostructure is shown in Fig.~\ref{Fig1}(a).
The MAT4G is fabricated into a multi-probe Hall bar, where electrostatic tuning is achieved using a gold top gate (TG), a gold finger gate (FG), and a graphite bottom gate (BG). This structure allows us to engineer a gate-defined JJ with a width $2W=\SI{1.1}{\upmu m}$ along $y$, a length $L = 12W$ along $x$ and a junction length $L_\textnormal{j}=\SI{150}{nm}$, as illustrated in Fig.~\ref{Fig1}(b). Beyond the device geometry, the graphene stacking layout plays a pivotal role; each successive layer in the MAT4G is rotated by an alternating angle of $\pm\theta\approx1.6^{\circ}$ relative to the previous one. This experimentally extracted angle is slightly below the theoretical magic-angle of $\theta_M=1.75^{\circ}$ \cite{khalaf2019magic}. Due to this alternating even-layer structure, the material lacks a horizontal mirror symmetry $\sigma_h$, as depicted in Fig.~\ref{Fig1}(c). Such symmetry exists exclusively in alternating twist-angle structures with an odd layer number (e.g., twisted trilayer graphene), whereas it is broken in structures with an even layer number (e.g., twisted bilayer graphene or our MAT4G). However, we note that these symmetry arguments strictly apply to ideal heterostructures, while in real samples the presence of moiré inhomogeneities may modify the local symmetry. 

The bulk phase diagram is shown in Fig.~\ref{Fig1}(d) (reproduced from Fig.~1 in Ref.~\cite{perego2024experimental}), where the four-terminal voltage across the device $V$ is recorded as a function of the lead carrier density and displacement field, $n_\textnormal{l}$ and $D_\textnormal{l}$, at fixed d.c.\ current $I=\SI{10}{nA}$ and temperature $T=\SI{50}{mK}$. Two distinct superconducting domes are observed, corresponding to the electron-doped and hole-doped regimes, where the voltage drops to zero (indicated by the dark blue regions), consistent with observations in other MAT4G realizations \cite{park2022robust,zhang2022promotion,burg2022emergence}. Unlike twisted bilayer graphene, superconductivity in MAT4G can be suppressed by applying a sufficiently large displacement field. In our device, we find that $D_\textnormal{l}/\epsilon_0\approx\SI{0.6}{V/nm}$ quenches superconductivity (where $\epsilon_0$ denotes the vacuum permittivity). The presence of extended superconducting regions in the phase diagram enables us to tune the leads of the JJ to different points of the superconducting domes, thereby varying the properties of the superconducting leads, including the critical field, critical temperature, critical current, superfluid stiffness, and gap magnitude (see Ref.~\cite{perego2024experimental} for further details). In this work, we carry out all measurements with the leads tuned to the centre of the electron-doped superconducting dome, highlighted by the orange rhombus in Fig.~\ref{Fig1}(d) ($n_\textnormal{l} = \SI{4.2e12}{cm^{-2}}$, $D_\textnormal{l}/\epsilon_0 = \SI{-0.3}{V/nm}$). The junction barrier is instead tuned to the electron-doped full filling $\nu\approx 4$ (where $\nu$ is the filling factor denoting the number of charge carriers per moiré superlattice unit cell), highlighted by the red dot in Fig.~\ref{Fig1}(d). The junction density and displacement field are fixed at $n_\textnormal{j} = \SI{6.2e12}{cm^{-2}}$ and ${D_\textnormal{j}/\epsilon_0 = \SI{-0.5}{V/nm}}$, respectively. Electrostatic simulations using these tuning parameters confirm a junction length of $L_\textnormal{j}\approx\SI{150}{nm}$, consistent with fabrication constraints as highlighted in Fig.~\ref{Fig1}(e).

\subsection*{Josephson diode effect with out-of-plane field $B_z$}
\begin{figure}
\includegraphics[width=1\columnwidth]{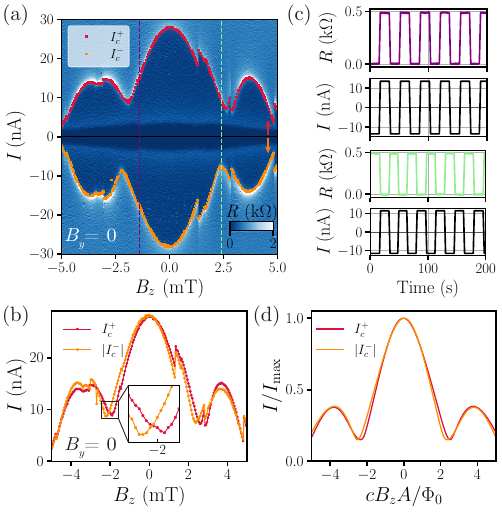}
\caption{(a) Differential resistance $R$ measured as a function of $I$ and $B_z$ at $B_y=0$ for the device tuned at $n_\textnormal{l} = \SI{4.2e12}{cm^{-2}}$, $D_\textnormal{l}/\epsilon_0 = \SI{-0.3}{V/nm}$, $n_\textnormal{j} = \SI{6.2e12}{cm^{-2}}$ and ${D_\textnormal{j}/\epsilon_0 = \SI{-0.5}{V/nm}}$ (orange-red-orange setting in Fig.~\ref{Fig1}(d)). The positive $I_{c}^{+}$ and negative $I_{c}^{-}$ critical currents are highlighted in red and orange, respectively. The red and orange arrows indicate that the current is swept from zero to positive/negative values. (b) Extracted $I_{c}^{+}$ and $|I_{c}^{-}|$ from (a), showing that they differ $I_{c}^{+}(B_z)\neq |I_{c}^{-}(B_z)|$ indicating the presence of the JDE. The inset shows a zoom-in of the data around $B_z=\SI{-2}{mT}$. (c) Demonstration of the JDE, evidenced by the transition between superconducting and normal states when the applied current is reversed. The reverse diode effect is shown in purple (measured at $B_z=\SI{-1.4}{mT}$, see dashed purple line in (a)), and the forward diode effect is shown in green (measured at $B_z=\SI{2.4}{mT}$, see dashed green line in (a)). (d) Simulation using the theoretical model discussed in the text at $B_y=0$. The critical currents $I_{c}^{+}$ and $|I_{c}^{-}|$ normalised to the maximum $I_\textrm{max}$ are shown in red and orange as a function of the out-of-plane field $B_z$. An asymmetry is observed in the vicinity of the first minima. 
}
\label{Fig2}
\end{figure}

By applying a perpendicular magnetic field $B_z$ to the junction, a Fraunhofer-like interference pattern (FP) is observed, as shown in Fig.~\ref{Fig2}(a). The resulting field modulation of the critical current $I_\mathrm{c}(B_z)$ is characteristic of a two-dimensional junction, where the pattern decays as $1/\sqrt{B_z}$, in contrast to the $1/B_z$ decay observed in three-dimensional cases \cite{Clem2010,Tinkham2004,perego2024experimental}. The sharp discontinuities observed in the FP in Fig.~\ref{Fig2}(a) are attributed to the entry or exit of Pearl vortices \cite{Pearl1964} in the superconducting leads, as discussed in \cite{perego2024experimental, perego2026pearl, Perego2026}. When a vortex enters the leads in proximity to the junction, it induces a reduction in the Josephson phase; this manifests as a discrete shift of the $I_\mathrm{c}(B_z)$ \cite{Clem2011, KoganMints2014, KoganMints_PC2014}. In the experimental tuning used in this work, vortex dynamics occur on a long timescale ($\sim \SIrange{e2}{e3}{s}$), a condition specifically chosen to minimise vortex motion within the measurement window. Nevertheless, sweeping the out-of-plane magnetic field $B_z$ induces stochastic vortex entry and exit, manifesting as discrete events separated by long, intermittent intervals. This stochastic motion during a single FP measurement implies that the device's vortex configuration is not static but dynamic. Consequently, the state of the device at the onset of the $B_z$ sweep for the FP data may differ significantly from its state during or after the measurement. Moreover, by measuring the FP from positive to negative $B_z$, vortices can get pinned and therefore be present even at $B_z=0$. The presence of a single vortex effectively lifts both time-reversal (TRS) and inversion/$C_{2z}$ symmetries, providing a mechanism for the manifestation of non-reciprocal effects (even at $B_z=0$), see similar work on Abrikosov vortices in Ref.~\cite{golod2022demonstration}.

Interestingly, in the FP measured in Fig.~\ref{Fig2}(a), we observe an asymmetry between the positive ($I_{c}^{+}$ in red) and negative ($I_{c}^{-}$ in orange) critical currents. We define the critical current as the current bias required to reach a voltage threshold of $\SI{0.3}{\upmu V}$ (see Supplementary Information for further details). For each value of $B_z$, both $I_{c}^{+}$ and $I_{c}^{-}$ are obtained by sweeping $I$ from zero to positive and negative values respectively; $B_z$ is then stepped and the measurement repeated. Sweeping the current $I$ from zero minimises self-induced heating effects.

To better visualise the observed diode-like behaviour, Fig.~\ref{Fig2}(b) plots $I_{c}^{+}$ and the absolute value $|I_{c}^{-}|$. A difference is observed between the two, i.e., $I_{c}^{+}(B_z)\neq |I_{c}^{-}(B_z)|$, suggesting the presence of the JDE.

In general, TRS in planar JJs is broken by applying an out-of-plane magnetic field $B_z$ or by the presence of a pinned vortex. Conversely, in our device, $C_{2z}$ symmetry is broken by the moiré termination at the junction edges (reflecting the junction's inhomogeneity along the $y$-direction, see Fig.~\ref{Fig1}(b)) or by pinned vortices in the leads. Furthermore, vortex dynamics during FP measurements intrinsically break TRS. Because this is a dynamic process, an asymmetry persists (even at $B_z=0$) as confirmed by the inequality $I_{c}^{+}(B_z)\neq |I_{c}^{-}(-B_z)|$. As evidenced in Fig.~\ref{Fig2}(b), the strongest asymmetry is recorded around the first FP minima, where the critical current difference $\Delta I_{c}= I_{c}^{+}(B_z) - |I_{c}^{-}(B_z)|$ is maximised, with a maximum absolute diode efficiency $|\eta|=|\frac{I_{c}^+-|I_{c}^-|}{I_{c}^++|I_{c}^-|}|\approx14\%$. Consequently, by fixing $B_z$ near a FP minimum, the device can operate as either a forward or reverse diode, as demonstrated in Fig.~\ref{Fig2}(c). For instance, by fixing $B_z$ near the first positive minimum (green dashed line in Fig.~\ref{Fig2}(a)), a forward diode effect is achieved; as the d.c.\ current bias varies between $\pm\SI{11}{nA}$ (black $I$ trace in Fig.~\ref{Fig2}(c)), the sample switches between the superconducting and normal states (green $R$ trace in Fig.~\ref{Fig2}(c)).

\subsection*{Theoretical model}
In the following, we discuss a theoretical picture for the observed JDE. The free-energy for the JJ sketched in Fig.~\ref{Fig1}(b) is given by
\begin{equation}
\mathcal{F}=\mathcal{F}_0+\gamma_1 \Delta_\mathrm{L}^* \Delta_\mathrm{R}^{\phantom{*}}+\frac{1}{2} \gamma_2\left(\Delta_\mathrm{L}^* \Delta_\mathrm{R}^{\phantom{*}}\right)^2+\text { c.c }+\ldots
\label{Eqn_free_energy}
\end{equation}
where $\mathcal{F_\mathrm{0}}$ is the free energy in the absence of the Josephson coupling, $\Delta_\mathrm{L,R}=|\Delta_\mathrm{L,R}|e^{i\phi_{\mathrm{L,R}}}$ are the superconducting order parameters for the left and right leads, while $\gamma_\mathrm{1,2}$ are the coupling coefficients between the two order parameters; we also include the quartic coupling term, $\propto \gamma_2$, which is necessary to capture non-reciprocal behaviour. The coefficients $\gamma_{1,2}$ are complex since both TRS and $C_{2z}$ symmetry are broken. We write $\gamma_\mathrm{1,2}=|\gamma_\mathrm{1,2}|e^{i\theta_{1,2}}$. Furthermore, the moiré nature of the junction material is expected to introduce significant spatial inhomogeneities along the $y$-axis, directly affecting the Josephson coupling parameters $\gamma_{1,2}$. While such variations are typically negligible in atomically sharp, non-moiré junctions, they are expected to become prominent in gate-defined moiré devices, especially considering twist-angle disorder \cite{RothsteinGateTunable,inhomogeneities_uri2020mapping}. To account for this $y$-dependence, we introduce inversion-symmetry-breaking form factors $f_{1,2}(y)  \neq f_{1,2}(-y)$, leading to $y$-dependent Josephson couplings, $\gamma_{1,2}\rightarrow \gamma_{1,2} f_{1,2}(y)$.

Generalizing Eq.~\eqref{Eqn_free_energy} to the laterally inhomogeneous case and including the impact of the magnetic field, the Josephson current-phase relation is obtained via $I(\phi)=\frac{2e}{\hbar}\frac{\partial\mathcal{F}}{\partial\phi}$ and reads as (assuming for simplicity $|\Delta_\mathrm{L}|=|\Delta_\mathrm{R}| \equiv |\Delta|$)
\begin{equation}
\begin{aligned}
I(\phi) &= -\frac{2e}{\hbar}|\Delta|^2 \int_{-W}^W dy \, \Big[ |\gamma_1| f_1(y) \sin \left( \phi + \Phi_B(y) \right) \\
&\qquad\,\,  + \frac{|\gamma_2|}{2} |\Delta|^2 f_2(y) \sin \left( 2\phi + \delta + 2\Phi_B(y) \right) \Big].
\end{aligned}
\label{Eqn_current}
\end{equation}
Here, $e$ is the electron charge, $\hbar$ is the reduced Planck constant, and $\delta=\theta_2-2\theta_1=:B_y/B_{y,0}$ encodes the dependence on the in-plane magnetic field $B_y$. We will discuss the physical origin of this term below.
The out-of-plane magnetic field-induced phase modulation for a two-dimensional superconductor is given by \cite{Clem2010}
\begin{equation}
\Phi_B(y, B_z) = 1.7 \frac{B_z A}{\Phi_0} \sin \left( \frac{\pi y}{2 W} \right),
\end{equation}
where $A=4W^2$ and $\Phi_0$ is the superconducting flux quantum. For further details on the theoretical model, see the Supplementary Information.

By evaluating Eq.~\eqref{Eqn_current} at $B_y = 0$ ($\delta=0)$, we calculate the $B_z$ dependence of the positive and negative critical currents, $I_{c}^{+}$ and $I_{c}^{-}$. The resulting profiles as a function of $B_z$ are presented in Fig.~\ref{Fig2}(d), illustrating the non-reciprocal transport induced by the device's broken symmetries. In particular, in the theoretical model, TRS is broken by $B_z$ while $C_{2z}$ symmetry is broken by the form factors $f_{1,2}(y)$. For moderately large inhomogeneities in $\gamma_{1,2}$, varying by 30\% and 20\%, respectively, we obtain a critical current profile that matches the main features of the experimental results, compare Fig.~\ref{Fig2}(b) and (d).
In both cases, the most pronounced asymmetry occurs in the vicinity of the FP minima, where the difference between $I_{c}^{+}$ and $|I_{c}^{-}|$ is maximised. Furthermore, the simulation accurately reproduces the non-vanishing critical current at these minima. This lifting of the nodes is a direct consequence of the spatial modulations defined by $f_{1,2}(y)$. Note that in our theoretical model, vortex dynamics is not taken into account, leading to the modelled device not intrinsically breaking TRS such that $I_{c}^{+}(B_z)= |I_{c}^{-}(-B_z)|$ in Fig.~\ref{Fig2}(d).

\subsection*{Josephson diode effect with in-plane field $B_y$}
\begin{figure}
\includegraphics[width=\columnwidth]{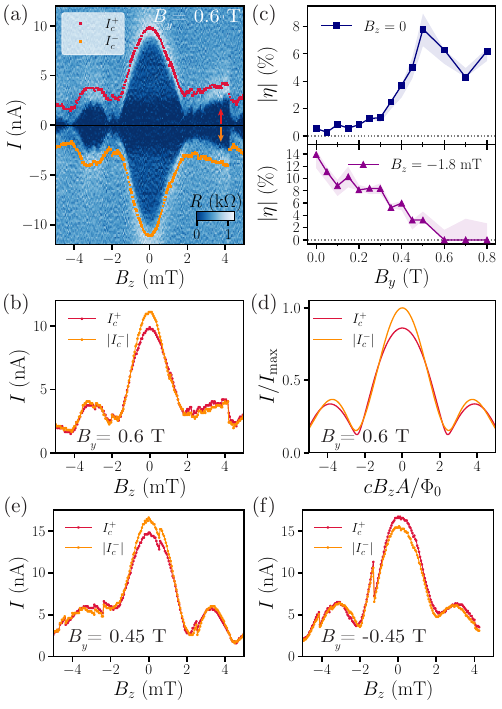}
\caption{(a) Differential resistance $R$ measured as a function of $I$ and $B_z$ at $B_y=\SI{0.6}{T}$ for the device tuned at $n_\textnormal{l} = \SI{4.2e12}{cm^{-2}}$, $D_\textnormal{l}/\epsilon_0 = \SI{-0.3}{V/nm}$, $n_\textnormal{j} = \SI{6.2e12}{cm^{-2}}$ and ${D_\textnormal{j}/\epsilon_0 = \SI{-0.5}{V/nm}}$ (orange-red-orange setting in Fig.~\ref{Fig1}(d)). The positive $I_{c}^{+}$ and negative $I_{c}^{-}$ critical currents are highlighted in red and orange, respectively. The red and orange arrows indicate that the current is swept from zero to positive/negative values. (b) Extracted $I_{c}^{+}$ and $|I_\mathrm{c}^{-}|$ from (a), showing the appearance of a JDE around zero out-of-plane field. 
(c) Absolute diode efficiency $|\eta|$ as a function of $B_y$ at $B_z = 0$ (top) and $B_z = \SI{-1.8}{mT}$ (bottom). Squares and triangles correspond to a nominal threshold for the critical current extraction of $V=\SI{0.5}{\upmu V}$. The shaded envelope bounds the spread obtained by varying the voltage threshold between $V=\SI{0.3}{\upmu V}$ and $V=\SI{1.0}{\upmu V}$. (d) Simulation using the theoretical model discussed in the text at $\delta=-1.5$. 
The critical currents $I_{c}^{+}$ and $|I_{c}^{-}|$ normalised to the maximum $I_c$ are shown in red and orange as a function of the out-of-plane field $B_z$. A strong asymmetry is observed around zero $B_z$. (e,f) Extracted $I_{c}^{+}$ and $|I_{c}^{-}|$ as a function of $B_z$ at $B_y = \SI{0.45}{T}$ (e) and $B_y = \SI{-0.45}{T}$ (f). Reversing the sign of the in-plane field inverts the polarity of the diode effect: at the central lobe, $|I_{c}^{-}| > I_{c}^{+}$ for positive $B_y$ (e), while $I_{c}^{+} > |I_{c}^{-}|$ for negative $B_y$ (f).
}%
\label{Fig3} 
\end{figure}

In the following section and in Fig.~\ref{Fig3}, we experimentally investigate the evolution of the JDE under a finite in-plane magnetic field, $B_y$.

Upon recording the FPs under an increasing $B_y$, two primary modifications emerge compared to the low-field regime. First, the $V-I$ characteristics across the JJ become smoother with no sharp jump at $I_{c}$ (see Supplementary Information for $V-I$ traces evolution).
Second, for fields exceeding $B_y \approx \SI{0.4}{T}$, a JDE becomes clearly visible at and near zero out-of-plane field ($B_z \approx 0$), as shown by the pronounced asymmetry between the extracted $I_{c}^{+}$ and $|I_{c}^{-}|$ in Fig.~\ref{Fig3}(a-b) at $B_y=\SI{0.6}{T}$ (measurements are corrected for a small misalignment angle between the device and the magnet, see Supplementary Information). The magnitude of this non-reciprocity increases with the applied in-plane field $B_y$. As shown by the critical currents extracted at $B_y = 0.6~\mathrm{T}$ (Fig.~\ref{Fig3}(b)), the asymmetry peaks near $B_z \approx 0$, yielding a maximum absolute diode efficiency of $|\eta| \approx 6\%$.
To track this evolution systematically, we extract the absolute diode efficiency $|\eta|$ as a function of $B_y$. Figure~\ref{Fig3}(c) summarises these results, comparing the behaviour at $B_z=0$ (top panel) with that near the first minimum of the FP at $B_z=\SI{-1.8}{mT}$ (bottom panel). At $B_z = 0$ (Fig.~\ref{Fig3}(c), top panel), the efficiency grows with the in-plane magnetic component, demonstrating the emergence of an in-plane mediated JDE. Interestingly, this diode effect rapidly vanishes as $B_z$ moves away from zero, as evidenced by the symmetric recovery of the secondary lobes in the FP in Fig.~\ref{Fig3}(a-b). In contrast to the out-of-plane zero-field behaviour, the efficiency near the FP minimum (Fig.~\ref{Fig3}(c), bottom panel) monotonically decreases with increasing $B_y$. This indicates that while the in-plane field induces a diode effect at the centre of the FP, it suppresses the asymmetry at finite $B_z$, effectively shifting the position of the maximum non-reciprocity from a finite out-of-plane field value at low $B_y$ toward zero out-of-plane field at high $B_y$.

Applying the theoretical framework outlined above, but now with a finite in-plane field, we calculate $I_{c}^{+}$ and $I_{c}^{-}$, which are shown in Fig.~\ref{Fig3}(d).
In the model, the in-plane field is incorporated via the parameter $\delta=B_y/B_{y,0}$, see following discussion section. Good agreement is found when associating $B_y = \SI{0.6}{T}$ with $\delta=-1.5$.
More importantly, upon turning on $\delta$ (or equivalently $B_y$), the simulation successfully reproduces the primary experimental feature: the most pronounced asymmetry, $\Delta I_{c}$, moves away from the vicinity of the first minima of the FP to being located near $B_z=0$.

Note that vortex dynamics are unlikely to explain the JDE that emerges with increasing $B_y$. Due to the ultrathin nature of MAT4G, its thickness $d\approx\SI{1}{nm}$ is significantly smaller than the superconducting coherence length $\xi\approx\SI{40}{nm}$ \cite{perego2024experimental} ($d\ll\xi$), implying that the formation of vortices induced by the in-plane field $B_y$ is precluded \cite{Tinkham2004, gennes1966superconductivity}. However, we must consider that applying $B_y$ could modify the free-energy landscape for Pearl vortices entering or exiting the leads along $y$ \cite{Stejic1994, perego2026pearl,Perego2026}, thereby affecting their dynamics. Considering that the in-plane critical field for MAT4G is significantly larger ($B_y^\text{critical}>\SI{2}{T}$) than the onset field for the diode effect ($B_y \approx \SI{0.4}{T}$), any field-induced suppression of the coherence length (and consequently of the superconducting gap magnitude) is negligible in this low-field regime $B_y< B_y^\text{critical}$. Given that these superconducting quantities govern the formulation of the free-energy and remain unaltered in this regime, the overall free-energy landscape should be unperturbed \cite{Stejic1994}. However, we cannot fully exclude that $B_y$ modifies the vortex free-energy landscape to some extent. We therefore rely on the observed phenomenology to disfavour a vortex interpretation.
As previously discussed, vortex motion occurs stochastically as $B_z$ is swept. Breaking of TRS and $C_{2z}$ symmetry around $B_z\approx0$ would require a vortex to remain pinned while $B_z$ is swept around the main lobe. This pinning would have to occur consistently for every FP recorded with increasing $B_y$, a scenario which is considered improbable. Instead, the smoothness of the non-reciprocal response (Fig.~\ref{Fig3}(c)) suggests an intrinsic, global symmetry-breaking mechanism (such as in-plane orbital coupling associated with Peierls phases) rather than an extrinsic discrete one (like a single trapped vortex). This is consistent with the behaviour under field reversal: as shown in Fig.~\ref{Fig3}(e-f), inverting the sign of the in-plane field from $B_y=\SI{0.45}{T}$ to $B_y=\SI{-0.45}{T}$ reverses the polarity of the diode effect while preserving its magnitude (see Supplementary Information for the efficiency data with negative $B_y$). Such a symmetric, sign-reversing response is expected for an intrinsic $B_y$-driven mechanism --- in our model the in-plane coupling enters through $\delta = B_y/B_{y,0}$, which is odd in $B_y$. 

\section*{Discussion}
Observing a JDE at a finite out-of-plane field $B_z$ and zero in-plane field $B_y$, can be understood via TRS breaking by $B_z$ combined with $C_{2z}$ symmetry breaking from junction inhomogeneity along $y$. In contrast, a JDE emerging around zero out-of-plane field ($B_z=0$) and driven by an increasing in-plane field $B_y$ requires further interpretation. Within our theoretical framework, an increasing $B_y$ elevates the parameter $\delta$ in Eq. \ref{Eqn_current}, which dictates the influence of the in-plane field on the transport non-reciprocity. Uncovering the physical origin of $\delta$ is therefore essential to understand the mechanism driving this diode effect. One prominent possibility is that $\delta$ stems from in-plane orbital coupling associated with Peierls phases in the interlayer hopping matrix elements. As $B_y$ increases, this orbital coupling grows, thereby enhancing $\delta$ and enabling the diode effect to manifest at zero out-of-plane field. Indeed, both our experimental data and theoretical model show that the maximum critical current asymmetry shifts from a finite out-of-plane value ($B_z\neq0$) to zero ($B_z=0$) upon increasing $B_y$ (see Supplementary Information). Attributing $\delta$ to strong in-plane orbital coupling is consistent with expectations for multi-layer systems like MAT4G \cite{antebi2022plane, mandal2023valley, margetis2024optical, vasilevskiy2026plane, vishwanathInplaneCoupling}; see the Supplementary Information for a discussion of the associated field scale $B_{y,0}$. Crucially, this distinguishes the observed behaviour from phenomena driven by spin-orbit coupling \cite{baumgartner2022supercurrent,davydova2022universal}, which remains intrinsically weak in this carbon-based system \cite{huertas2006spin,konschuh2010tight}. Furthermore, due to the absence of horizontal mirror symmetry $\sigma_h$ (Fig.~\ref{Fig1}(b)), this in-plane orbital coupling is expected to be substantially stronger in even-layer twisted structures than in their odd-layer counterparts. This fundamental difference possibly explains why the in-plane $B_y$ induced SDE is absent in bulk twisted trilayer graphene \cite{lin2022zero}, whereas we observe an effect in our four-layer system, see Supplementary Information. Interestingly, we observe a Pauli limit violation (see Supplementary Information), as previously reported in bulk MAT4G \cite{park2022robust}. We note, however, that recent theoretical work predicts a weak in-plane orbital coupling for four-layer systems compared to bilayers at the first magic angle \cite{vasilevskiy2026plane}, contrasting our experimental observations. We propose that this discrepancy stems from the fact that these calculations are restricted to the MAT4G tuned at charge neutrality. In our study, the device is electrostatically tuned to a finite carrier density and displacement field, the latter strongly modifies the band structure of MAT4G, and possibly interaction effects could lead to additional contributions.

In summary, we have explored the SDE in a gate-defined MAT4G JJ under both out-of-plane and in-plane magnetic fields. While an out-of-plane field induces a JDE, reflecting the strong influence of the moiré nature on the system's symmetries, we also demonstrate the emergence of a JDE driven entirely by an in-plane field. We propose strong in-plane orbital coupling in MAT4G as a viable explanation for the observed SDE at zero out-of-plane field with increasing in-plane field. Our findings highlight the potential role of in-plane orbital effects in these twisted systems and suggest a high sensitivity to the number of twisted layers.

\section*{Methods}
\subsection*{Fabrication details}
We fabricated a MAT4G stack using the dry pick-up method~\cite{kim_vdw_2016}. All the details for the fabrication and tuning of this sample can be found in Ref.~\cite{perego2024experimental}.

\subsection*{Measurement setup}
All measurements were carried out in a dilution refrigerator with a base temperature of $\SI{7}{mK}$ unless stated otherwise. Our measurements are current biased, i.e., we apply a current and measure the voltage drop in a two-terminal configuration (we correct for contact resistances). To generate the bias current, we use a home-built d.c.\ voltage source in series with a $\SI{10}{M\Omega}$ or $\SI{100}{M\Omega}$ resistor. The measured voltage is amplified using a home-built low-noise d.c.\ amplifier (see Ref.~\cite{marki2017temperature}) and its output is measured with a Hewlett Packard 3441A digital multimeter. The bottom, top, and finger gates are connected to home-built low-noise d.c.\ voltage sources.

\section*{Data availability}
The data supporting the findings of this study, together with the code for plotting the figures, is available online through the ETH Research Collection at https://doi.org/10.3929/ethz-c-000801690.

\begin{acknowledgments}
We thank Peter M\"{a}rki and the staff of the ETH cleanroom facility FIRST for technical support. Financial support was provided by the European Graphene Flagship Core3 Project, H2020 European Research Council (ERC) Synergy Grant under Grant Agreement 951541, the European Union’s Horizon 2020 research and innovation program under grant agreement number 862660/QUANTUM E LEAPS, the European Innovation Council under grant agreement number 101046231/FantastiCOF, the EU Cost Action CA21144 (SUPERQUMAP). K.W. and T.T. acknowledge support from the JSPS KAKENHI (Grant Numbers 21H05233 and 23H02052) and the World Premier International Research Center Initiative (WPI), MEXT, Japan. C.G.A. acknowledges support from the Heidi Ras foundation via an ETH Quantum Center fellowship. 
S.B.~and M.S.S.~further acknowledge funding by the European Union (ERC-2021-STG, Project 101040651---SuperCorr). Views and opinions expressed are however those of the authors only and do not necessarily reflect those of the European Union or the European Research Council Executive Agency. Neither the European Union nor the granting authority can be held responsible for them.
\end{acknowledgments}

\section*{Author contributions}
M.P. fabricated the device. T.T. and K.W. supplied the hBN crystals. M.P. and M.A.E.H. performed the measurements. M.P. and M.A.E.H. analysed the data. S.B. and M.S.S. developed the theoretical model. G.Z. developed the COMSOL simulation. M.P. and S.B. wrote the manuscript, and all authors were involved in the reviewing process. M.P., M.A.E.H., C.G.A. and A.M.T. discussed the data. M.P. and M.A.E.H. conceived and designed the experiment. T.I. and K.E. supervised the work.

\section*{Competing Interests Statement}
The authors declare no competing interests.

\onecolumngrid 
\addcontentsline{toc}{section}{References}

\section*{\label{sec:refs}References}

\def\bibsection{} 
\bibliography{Bibliography}
\end{document}